\documentclass[manuscript]{aastex}

\shorttitle{Particle acceleration related to plasmoid motions}
\shortauthors{Takasao et al.}

\usepackage{bm}
\usepackage{amsmath}
\usepackage{color}
\usepackage{url}

\begin{document}

\title{OBSERVATIONAL EVIDENCE OF PARTICLE ACCELERATION ASSOCIATED WITH PLASMOID MOTIONS}

\author{Shinsuke Takasao\altaffilmark{1}}
\email{takasao@nagoya-u.jp}
\author{Ayumi Asai\altaffilmark{2,3}}
\author{Hiroaki Isobe\altaffilmark{3,4}}

\and

\author{Kazunari Shibata\altaffilmark{2,3}}

\altaffiltext{1}{Department of Physics, Nagoya University, Nagoya, Aichi 464-8602, Japan}
\altaffiltext{2}{Kwasan and Hida Observatories, Kyoto University, Yamashina, Kyoto 607-8471, Japan}
\altaffiltext{3}{Unit of Synergetic Studies for Space, Kyoto University, Yamashina, Kyoto 607-8471, Japan}
\altaffiltext{4}{Graduate School of Advanced Integrated Studies in Human Survivability, Kyoto University, 1 Yoshida Nakaadachi-cho, Sakyo-ku, Kyoto 606-8306, Japan}

\begin{abstract}
We report a strong association between the particle acceleration and plasma motions found in the 2010 August 18 solar flare. The plasma motions are tracked in the extreme-ultraviolet (EUV) images taken by the Atmospheric Imaging Assembly (AIA) on board the {\it Solar Dynamics Observatory} and the Extreme UltraViolet Imager (EUVI) on the {\it Solar Terrestrial Relation Observatory} spacecraft {\it Ahead}, and the signature of particle acceleration was investigated by using Nobeyama Radioheliograph data. In our previous paper, we reported that in EUV images many plasma blobs appeared in the current sheet above the flare arcade. They were ejected bidirectionally along the current sheet, and the blobs that were ejected sunward collided with the flare arcade. Some of them collided or merged with each other before they were ejected from the current sheet. We discovered impulsive radio bursts associated with such plasma motions (ejection, coalescence, and collision with the post flare loops). The radio bursts are considered to be the gyrosynchrotron radiation by nonthermal high energy electrons. In addition, the stereoscopic observation by AIA and EUVI suggests that plasma blobs had a three-dimensionally elongated structure. We consider that the plasma blobs were three-dimensional plasmoids (i.e. flux ropes) moving in a current sheet. We believe that our observation provides clear evidence of particle acceleration associated with the plasmoid motions. We discuss possible acceleration mechanisms on the basis of our results.
\end{abstract}


\section{INTRODUCTION}
Solar flares are the most energetic phenomenon in the solar system, and they rapidly convert a huge amount of magnetic energy (10$^{29}$-10$^{32}$~erg) into the thermal, kinetic, and nonthermal high energy particle energies \citep{shibata2011,fletcher2011,lin2015}. The central engine is believed to be magnetic reconnection, a physical process in which a magnetic field in a highly conducting plasma changes its topology due to finite resistivity \citep{priest2000}, though its physics has not been resolved yet. 

High energy electrons with the energy of 100~keV--1~MeV produce both hard X-ray and microwave emissions \citep{bastian1998}. Nonthermal hard X-ray emissions often show rapid subsecond time variability during the impulsive phase, and it is known that microwave radio emissions exhibit a similar temporal behavior \citep[e.g.][]{altyntsev2000,asai2004}. Therefore, it has been inferred that rapidly varying radio bursts are produced by the nonthermal high energy electrons. The emission mechanism of the radio bursts is believed to be gyrosynchrotron radiation, and the gyrosynchrotron radiation has been commonly studied approximately in the 1--30~GHz frequency range \citep{dulk1985}.

Hard X-ray and radio observations have revealed that a large number of particles are accelerated during flares, though the acceleration mechanisms still remain puzzling \citep{miller1997,aschwanden2002,krucker2010}. The acceleration site is generally assumed in the corona, and many observations suggest that particles seem to be accelerated in the reconnection outflows and/or reconnection regions \citep[e.g.][]{sui2003,liu2013}. Shocks in the outflows may also contribute to the acceleration \citep{tsuneta1998}. Therefore, investigating the outflow structure in detail is important for understanding of the particle acceleration mechanisms.

Recent observations and theories suggest that current sheets and reconnection outflows of solar flares are very dynamic with multiple plasmoids (regions of magnetically confined plasma). {\it Yohkoh} soft X-ray observations revealed a hot plasmoid ejections in the early phase of flares \citep{shibata1995,tsuneta1997,ohyama1998}. Radio spectrograph observations have shown drifting pulsating structures (DPS) in the 0.8--4.5~GHz frequency range \citep{karlicky2004}. The emission mechanism is considered to be the plasma emission generated by electron beams, which are injected into the plasmoids. It is considered that they are caused by plasmoids moving in a current sheet \citep[e.g.][]{karlicky2007}. Theoretical analyses and numerical simulations indicate that multiple plasmoids will be formed in the current sheet with a large ($>10^4$) Lundquist number like a coronal current sheet \citep{loureiro2007,shibata2001,bhattacharjee2009,barta2011}. Such reconnection is called ``plasmoid-dominated reconnection."

In our previous paper \citep{takasao2012}, we discovered many fine scale ($\sim3^{\prime\prime}$) plasma blobs in a current sheet of a flare using extreme-ultraviolet (EUV) images, and considered them to be plasmoids. We also reported the dynamical emergence, coalescence, and ejections of plasmoids, which are consistent with the plasmoid-dominated reconnection picture. Similar plasma blobs were also observed in other events \citep{liu2013,kumar2013}, and some fraction of the so-called supra-arcade downflows may also be plasmoids \citep{mckenzie1999,savage2010}. A recent comparison of spectroscopic observations and numerical simulations also suggests the formation of multiple plasmoids during flares \citep{innes2015}.

Theories suggest that plasmoids can play significant roles in accelerating particles. It has been pointed out that particles trapped in plasmoids can be efficiently accelerated by the curvature drift \citep{kliem1994}, by the secondary reconnection at the merging points \citep{oka2010}, by a Fermi acceleration due to the contraction of the plasmoids \citep{drake2006}, and by a Fermi acceleration through the interaction with fast shocks \citep{nishizuka2013}. The stochastic acceleration in the plasmoid-dominated reconnection is also studied \citep{lazarian2009,hoshino2012}.


There are few imaging observations to clarify the relation between the plasmoid motions and electron acceleration. It has been pointed out that dark supra-arcade downflows are associated with nonthermal emissions \citep{asai2004,nishizuka2010}. Many DPSs are also accompanied by hard X-ray emissions \citep{nishizuka2015}. \citet{milligan2010} found that the collision of a slowly moving ($\sim12$~km~s$^{-1}$), few 10$^{\prime\prime}$-scale X-ray source (hypothesized as a plasmoid) in {\it Reuven Ramaty High Energy Solar Spectroscopic Imager} \citep[{\it RHESSI};][]{lin2002} images leads to a gradual increase of nonthermal emissions (the increase in the timescale is a few minutes). These observations suggest the possibility that the plasmoid motion caused electron acceleration. However, the relation between the plasmoid motions and impulsive nonthermal emission bursts with a much shorter timescale has not been clarified yet. Thus, we need higher temporal and spatial resolution observations for a complete understanding of plasmoid dynamics associated with electron acceleration.

With the aim of investigating the relation between the plasmoid motions and electron acceleration, we studied the limb flare that occurred on 2010 August 18 using EUV and microwave radio images with high temporal and spatial resolutions. The overall characteristics of this flare were reported in our previous paper \citep{takasao2012}. We found successive formations and ejections of plasma blobs or plasmoids, and their interaction in the current sheet. We discovered several nonthermal radio bursts associated with such plasmoid motions. We will show that the locations and timings of those bursts are in good agreement with the plasmoid motions. This study indicates a strong association between the plasmoid motions and particle acceleration, and suggests possible acceleration processes.

\section{GLOBAL EVOLUTION OF THIS FLARE}
We analyzed data of the flare (C4.5 on the {\it GOES} class) that occurred in the Active Region NOAA~11099 on 2010 August 18. The active region was located beyond the northwest limb at the time when the flare occurred. This flare was observed by the Extreme UltraViolet Imager \citep[EUVI;][]{howard2008,wuelser2004} on the {\it Solar Terrestrial Relation Observatory} spacecraft {\it Ahead} \citep[STEREO-{\it A};][]{kaiser2008}, the Atmospheric Imaging Assembly (AIA) on the {\it Solar Dynamics Observatory (SDO)}, and Nobeyama Radioheliograph \citep[NoRH;][]{nakajima1994}. In this paper, we used 195~{\AA} images of EUVI, 193~{\AA} and 131~{\AA} images of AIA, and 34~GHz images of NoRH. At this radio frequency, thermal gyroresonance radiation from the chromosphere and nonthermal gyrosynchrotron radiation from accelerated electrons are expected.

The pixel size and the temporal resolution of the 195~{\AA} images obtained by EUVI are about 1$^{\prime\prime}$.6 and 150~s, respectively. Those of each wavelength images obtained by AIA are about 0.6$^{\prime\prime}$ and 12~s, respectively. The 195~{\AA} channel of EUVI contains Fe~{\sc xii} line with the formation temperature of 1.6~MK and Fe~{\sc xxiv} line with that of 20~MK. The narrowband EUV passbands of AIA are sensitive to different ionization states of iron. The 193~{\AA} channel is similar to the 195~{\AA} channel of EUVI. In this study the 193~{\AA} images are mostly used to track the plasma motions in the flaring region. In the 131~{\AA} channel Fe~{\sc xxi} line, formed at 11~MK, is dominant in flaring regions, while it also contains lower temperature lines such as Fe~{\sc viii} line, formed at 0.4~MK. The 131~{\AA} images are used to show the hot flare loops and hot plasma motions. 

The spatial resolution of NoRH 34~GHz images are about 10$^{\prime\prime}$. This spatial resolution is lower than the best value (5$^{\prime\prime}$) because of the elevation angle. We analyzed 1-sec integrated data, though the original temporal resolution is 0.1~sec. We used a standard CLEAN algorithm developed by Hanaoka (\url{http://solar.nro.nao.ac.jp/norh/soft/synthesis/hanaoka/}). The intensity of the radio emission is proportional to the brightness temperature. The brightness temperature is normalized so that the spatially averaged brightness temperature of the thermal emission from the disk (thermal gyroresonance radiation) is equal to the typical chromospheric temperature of 10$^4$~K. The high time cadence observations by NoRH enable us to study highly intermittent nonthermal emissions associated with the plasmoid motions. AIA and NoRH data were co-aligned following the information in the FITS header of data. We confirmed that the solar limbs in both images are matched.

\begin{figure}
\epsscale{1.0}
\plotone{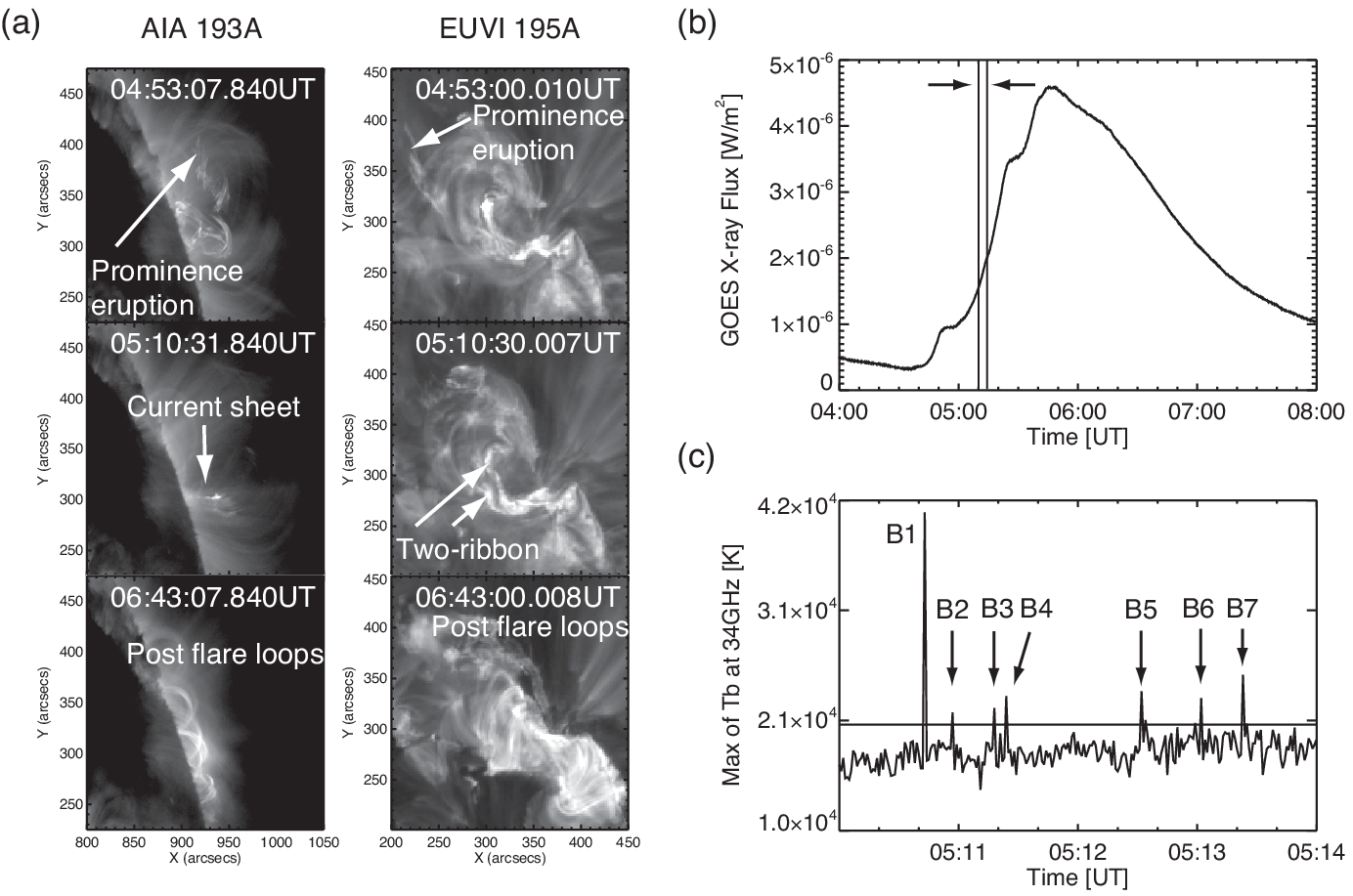}
\caption{Evolution of the 2010 August 18 flare. (a) EUV images taken by {\it SDO}/AIA and STEREO-{\it A}/EUVI. (b) {\it GOES} soft X-ray flux in the 1.0--8.0~{\AA} channel. (c) Temporal evolution of the maximum brightness temperature $T_b$ at 34~GHz measured in the flaring region. The radio bursts, interpreted as a signature of nonthermal emissions, are labeled as B1, B2,... and B7 (B denotes ``burst"). The horizontal line indicates the 3-$\sigma$ level ($2.0\times 10^4$~K). The duration indicated in the panel~(b) corresponds to the time span of the panel~(c). \label{fig:intro}}
\end{figure}

Figure~\ref{fig:intro}(a) shows the global evolution of this flare. We can obtain the side view of this flare from the AIA images, and the top view from the EUVI images ({\it STEREO-A}/EUVI was located ahead of the Earth with the separation angle of 79$^\circ$.877 at 05:00 UT). Before the flare, a prominence eruption occurred around 4:40 UT. Then, the soft X-ray flux showed a rapid increase (Figure~\ref{fig:intro}(b)). EUVI images display that the flare showed a typical two-ribbon structure in the early phase, and then an elongated flare arcade in the decay phase. We also note that this flare was associated with a coronal mass ejection.

\begin{figure}
\epsscale{.90}
\plotone{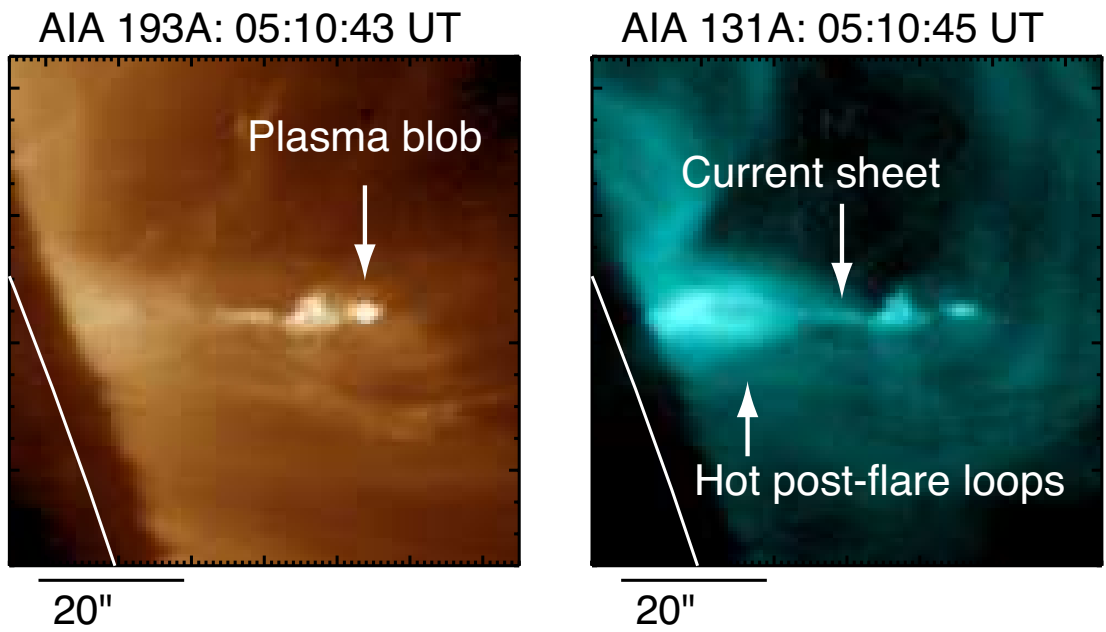}
\caption{Close-up images of the reconnection site in 193~{\AA} (Left) and 131~{\AA} (Right) of AIA at the time when the current sheet, plasmoid, and hot flare loops are observed. White solid lines denote the solar limb.  \label{fig:aia_intro}}
\end{figure}

As reported by \citet{takasao2012}, a bright straight structure appeared above the flare loops during the time between 05:10-05:14 UT (Figure~\ref{fig:aia_intro}). The straight structure was persistent in AIA 131~{\AA} images throughout the duration. As we will show later, AIA 193~{\AA} images show that many plasma blobs were formed and ejected bidirectionally from it. The reconnection inflow toward the straight structure was also observed. Since the overall characteristics of this flare are consistent with the classical CSHKP model \citep{carmichael1964,sturrock1966,hirayama1974,kopp1976}, we consider that the straight structure was the current sheet and the plasma blobs are plasmoids.

We examined the signatures of nonthermal emissions using NoRH data when the current sheet was visible in AIA images. Figure~\ref{fig:intro}(c) exhibits the temporal evolution of the maximum brightness temperature $T_b$~[K] at 34~GHz in the flaring region (we used the region of $[890^{\prime\prime},970^{\prime\prime}]\times[275^{\prime\prime},335^{\prime\prime}]$ of the heliospheric coordinates. We focus on the plasmoid motions in this flaring region). We found several intermittent radio bursts (the radio bursts introduced in this paper are indicated by B1, B2,..., and B7. B denotes ``burst"). We note that the 3$\sigma$ level is $T_b=2.0\times10^4$~K, and these bursts are detected above this level. Thus, these bursts are distinguishable from noise spikes. Radiation cooling of the thermal plasma cannot explain this rapid disappearance of the high $T_b$ regions. Thus, we consider that the radio bursts are gyrosynchrotron radiation by nonthermal high energy electrons. In the rest of the paper, we will show the relation between these radio bursts and plasma blob motions using NoRH 34~GHz and AIA 193 and 131~{\AA} images. {\it RHESSI} did not observe the flare during the impulsive phase, because it was {\it RHESSI} night time.

\section{RADIO BURSTS ASSOCIATED WITH PLASMA BLOB MOTIONS}
\subsection{Examples of Ejection, Coalescence, and Collision with Flare Loops}
During the rising phase of this flare, many plasmoids appeared, and then were ejected or collided with each other in the current sheet above the flare arcade. We investigated the relation between such plasmoid motions and occurrence of the radio bursts. In the following, we estimate the speeds of plasmoids. The speeds should contain an error of $d_p/\Delta t_p\simeq 40--50$~km~s$^{-1}$, where $d_p$ is the typical size of the plasmoids ($\sim3^{\prime\prime}$), and $\Delta t_p$ is the typical timescale of the plasmoid motions ($\sim$50~s).

\begin{figure}
\epsscale{.70}
\plotone{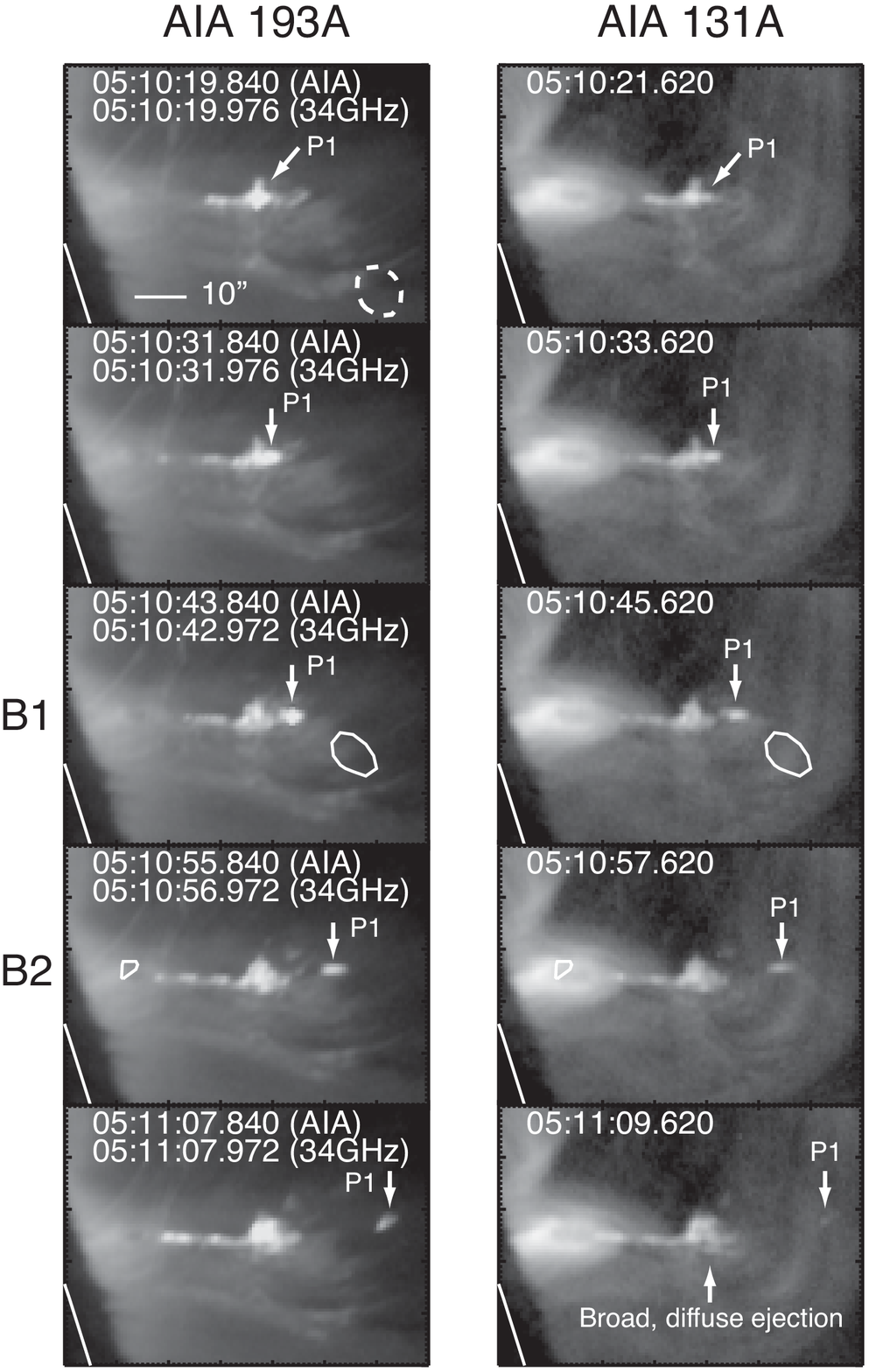}
\caption{Radio bursts B1 and B2. Left: AIA 193~{\AA} images. Right: AIA 131~{\AA} images. All images are overplotted with the NoRH 34~GHz 3-$\sigma$ contours ($2.0\times10^4$~K). All the times shown are UT. The dashed contour line shows the beam size. \label{fig:burst12}}
\end{figure}

Figure~\ref{fig:burst12} shows the example of the the plasmoid ejected upward from the current sheet. The plasmoid P1 is indicated by arrows. The apparent velocity is $\sim$400~km~s$^{-1}$. We note that the ejection of P1 was followed by a broad, diffuse plasma ejection whose velocity is $\sim$220~km~s$^{-1}$. We also see the downward elongation of the straight structure in 193~{\AA} images. The radio burst B1 was detected just after the ejection of P1. The radio emission site is separated from the ejection site by approximately 20$^{\prime\prime}$. This separation distance is larger than the spatial resolution, but we note the match of the timing of the burst B1 and the plasmoid ejection. The radio burst B2 was detected when the straight structure in 193~{\AA} images extended to the hot flare arcade seen in 131~{\AA} images. The spatial and temporal resolutions of AIA are insufficient to study the plasma motions related to the burst B2.

\begin{figure}
\epsscale{.80}
\plotone{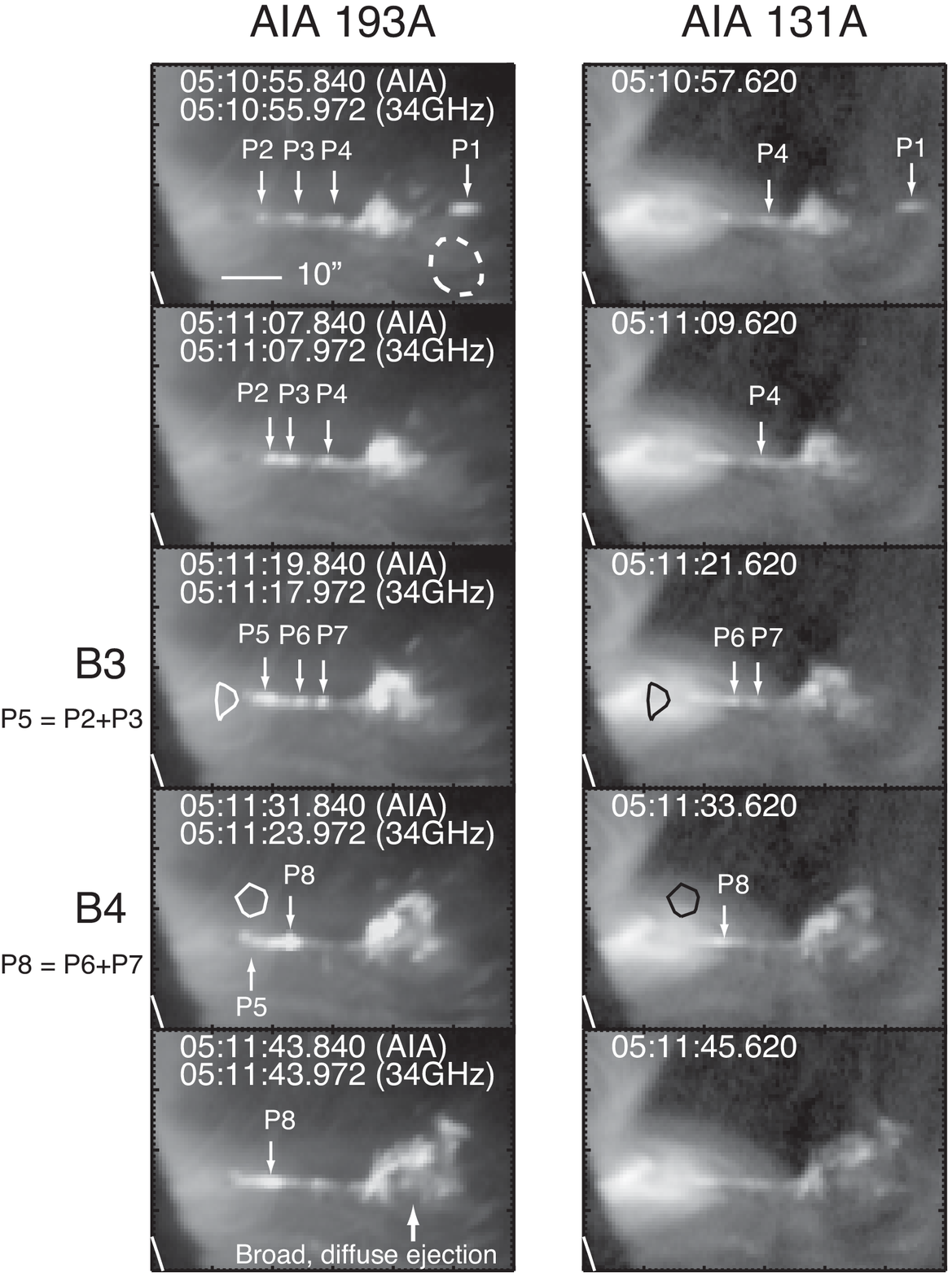}
\caption{Radio bursts B3 and B4. Left: AIA 193~{\AA} images. Right: AIA 131~{\AA} images. All images are overplotted with the NoRH 34~GHz 3-$\sigma$ contours ($2.0\times10^4$~K). All the times shown are UT. The dashed contour line shows the beam size.\label{fig:burst34}}
\end{figure}

Many plasmoids successively appeared in the current sheet. Figure~\ref{fig:burst34} displays the downward ejections and possibly the coalescence of plasmoids. In the first panel in 193~{\AA} images, three plasmoids (P2, P3, and P4) are recognized in the current sheet. They became bigger, and P3 simultaneously moved slightly downward in 12~s. Considering the downward motion of P3, we assume that the large blob P5 at 05:11:19.840 UT was formed by the coalescence between P2 and P3. The coalescence of P2 and P3 will reduce the number of plasmoids from three (P2, P3, and P4) to two, but at 05:11:19.840~UT we can still find three plasmoids (P5, P6, and P7). We interpret this as an indication that either P6 and P7 is P4 and the other is a newly created plasmoid (the current temporal resolution is insufficient to distinguish these two). The fourth panel of 193~{\AA} image shows that P5 was ejected downward with the speed of $\sim250$~km~s$^{-1}$ and collided with the flare arcade. The downward speed was largely decelerated after the collision (the speed became $\sim$80~km~s$^{-1}$), and the apparent structure became diffuse. P6 and P7 collided and possibly merged with each other to form the lager blob P8. P8 was quickly ejected downward. The radio burst B3 was detected at around the time when P2 and P3 collided and possibly merged with each other to form P5. The radio burst B4 was observed at around the time when P5 collided with the flare arcade, and when P6 and P7 collided and possibly merged with each other to form P8. The time cadence of AIA (12~s) is not enough to conclude that the burst B4 was caused by either the former or the latter event.

\begin{figure}
\epsscale{.80}
\plotone{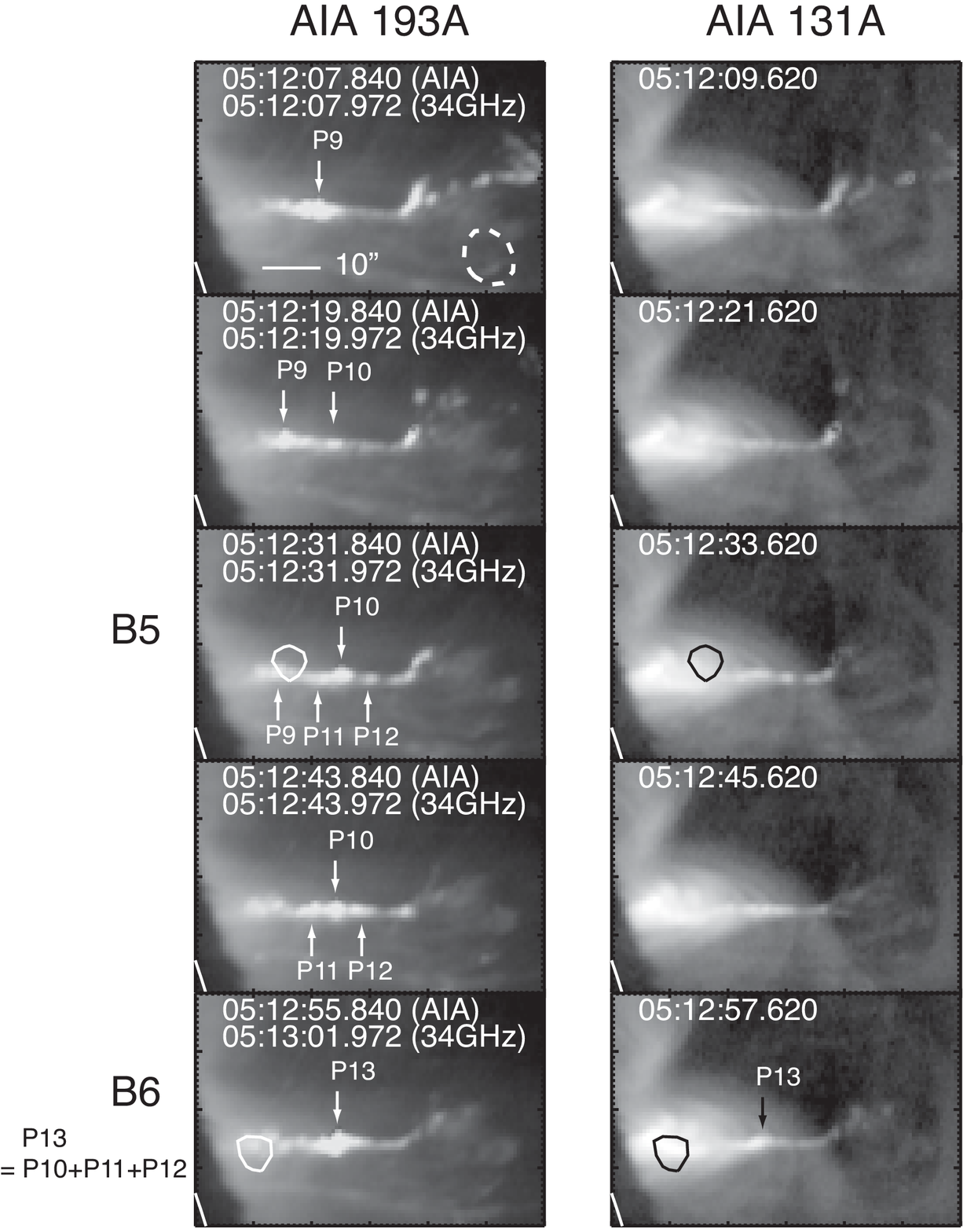}
\caption{Radio bursts B5 and B6. Left: AIA 193~{\AA} images. Right: AIA 131~{\AA} images. All images are overplotted with the NoRH 34~GHz 3-$\sigma$ contours ($2.0\times10^4$~K). All the times shown are UT. The dashed contour line shows the beam size.\label{fig:burst56}}
\end{figure}

Figure~\ref{fig:burst56} displays the plasmoid motions associated with the radio bursts B5 and B6. The radio burst B5 was detected when the plasmoid P9 seemed to collide with the flare arcade. The apparent structure of P9 became diffuse after the collision. The location and timing of the radio burst coincide well with those of the collision. At 05:12:31.840~UT (the third row of Figure~\ref{fig:burst56}), three plasmoids P10, P11, and P12 closely sit in the current sheet, and became larger with time (05:12:43.840~UT; the third row of Figure~\ref{fig:burst56}). They finally collided and possibly merged with each other to form the big plasmoid P13 at around 05:12:55.840~UT. The radio burst B6 was observed at this time, though the separation between the radio source and the coalescence site is larger than 10$^{\prime\prime}$.

\begin{figure}
\epsscale{.70}
\plotone{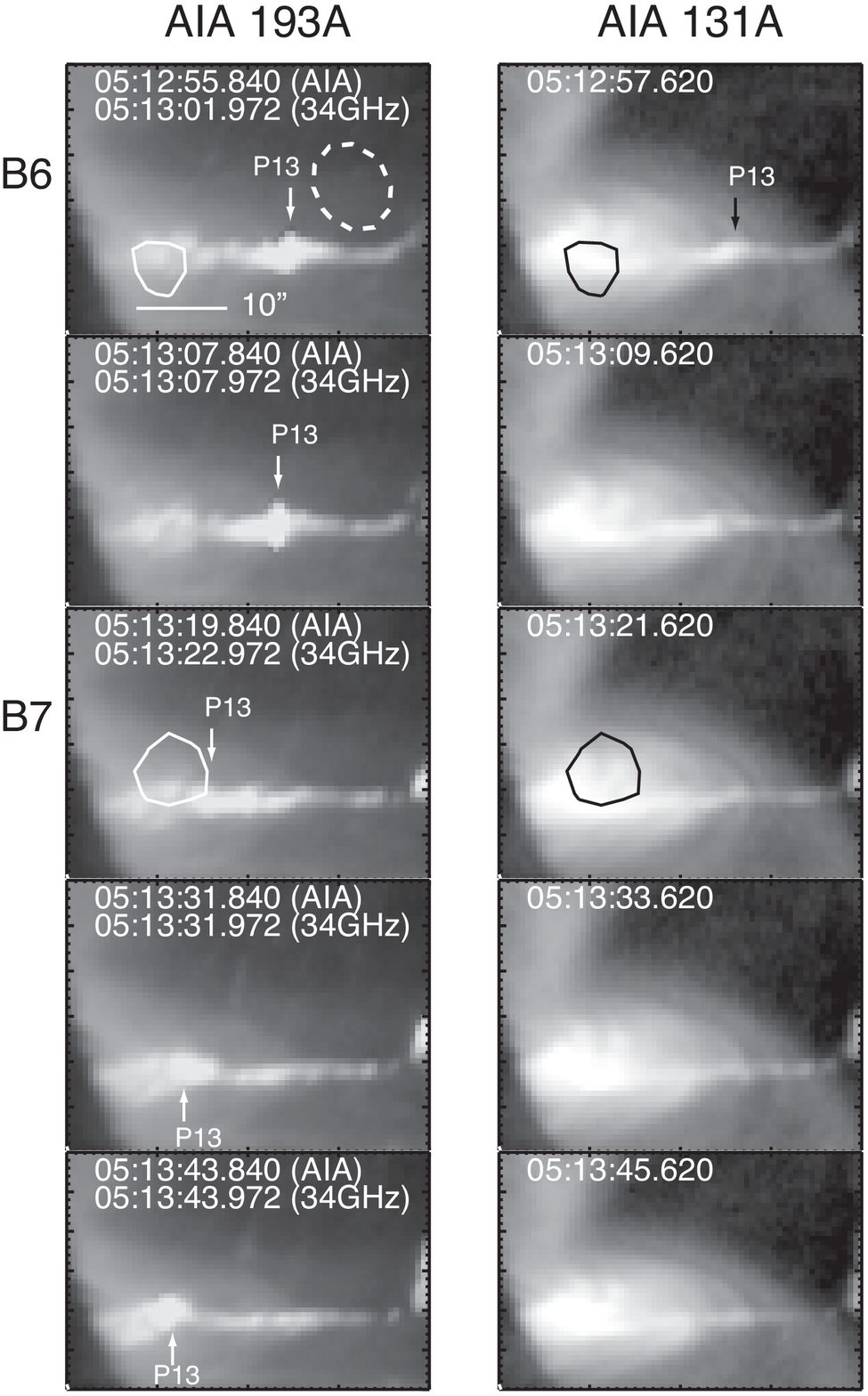}
\caption{Radio bursts B6 and B7. Left: AIA 193~{\AA} images. Right: AIA 131~{\AA} images. All images are overplotted with the NoRH 34~GHz 3-$\sigma$ contours ($2.0\times10^4$~K). All the times shown are UT. The dashed contour line shows the beam size.\label{fig:burst67}}
\end{figure}

Figure~\ref{fig:burst67} focused on the radio burst B7, which is related to the motion of the large ($\sim5^{\prime\prime}$) plasmoid P13. P13 moved downward with the speed of $\sim$280~km~s$^{-1}$. P13 eventually collided with the flare arcade and showed a similarly strong deceleration as other colliding plasmoids. The large radio source B7 was detected around the time when P13 collided with the flare arcade. The location and timing of the radio burst coincide well with those of the collision.

Figure~\ref{fig:slit_burst} highlights the association between the plasmoid motions and electron acceleration mentioned above. The plasma motions along the current sheet were investigated by using slits EW (E: east, W: west) shown in the left panel (we used two slits to track the plasma motions in the moving current sheet). The top-right panels show the time-sequenced images obtained along the slits. The intensity is averaged over the two slits. The solid lines track the apparent motion of the plasma. This figure indicates the bi-directional ejection, coalescence of plasmoids, and collision of plasmoids to the flare loops. The apparent velocities of the upward ejection of the plasmoid P1 and the broad diffuse ejection are 460~km~s$^{-1}$ and 220--250~km~s$^{-1}$, respectively. The downflow speed is 250--280~km~s$^{-1}$. We note that the downflows slowed down to approximately 70--80~km~s$^{-1}$ and spread out after colliding with the flare loops. The bottom-right panels of the figure show the temporal evolution of the maximum brightness temperature $T_b$ at 34~GHz (see also Figure~\ref{fig:intro}). The vertical dashed lines denote the timings of the plasmoid motions mentioned above, indicating a good match of the timing of the plasmoid motions (ejection, coalescence, and collision with the flare loops) and the radio bursts.

\subsection{Three-dimensional Structure of the Flare}
\begin{figure}
\epsscale{.90}
\plotone{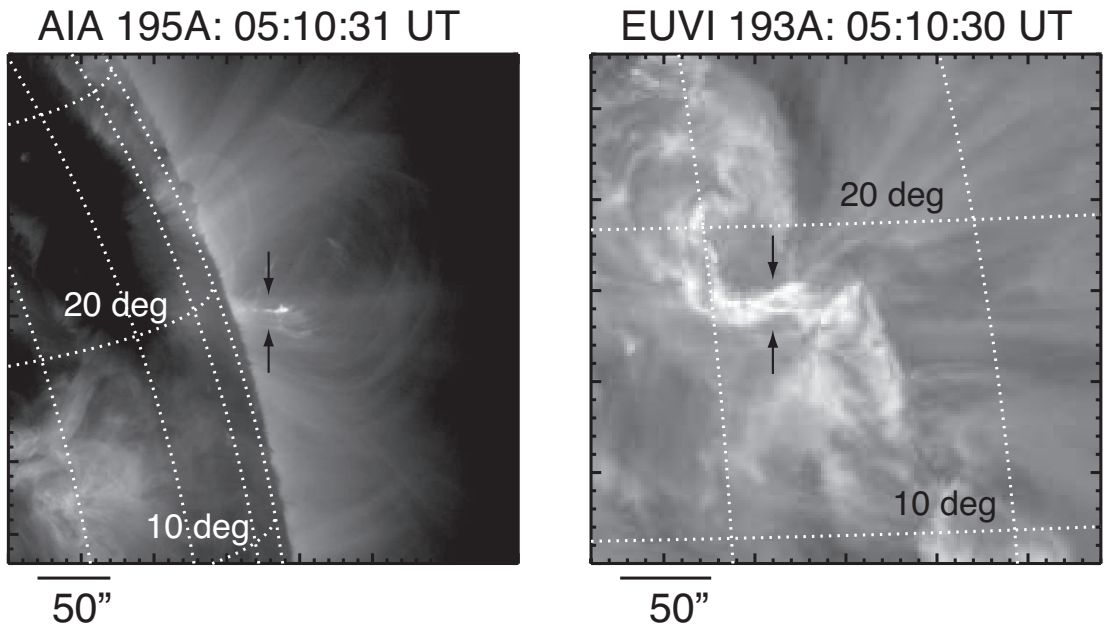}
\caption{The flare structure seen from the SDO view point (side-view, Right) and from the STEREO view point (top-view, Left) at the time when a plasmoid was ejected upward. Constant heliographic longitude and latitude are overplotted (grid spacing is 10~degree). Degrees of latitude are denoted. The bright straight structure in AIA 193~{\AA} is indicated by arrows. The brightening in EUVI 195~{\AA} is also indicated by arrows (see also Figure~\ref{fig:multi3d}). \label{fig:one3d}}
\end{figure}

Simultaneous observations by AIA and EUVI enabled us to study the three-dimensional (3D) structure of this flare. Figure~\ref{fig:one3d} displays the flare structure seen in two similar lines, namely AIA 193~{\AA} and in EUVI 195~{\AA}, at the time when the plasmoid P1 was ejected upward. Constant heliographic longitude and latitude are overplotted (grid spacing is 10~degree). The current sheet observed by AIA and the most bright region by EUVI are indicated by arrows. The latitudes of both structures are in good agreement. 

\begin{figure}
\epsscale{.90}
\plotone{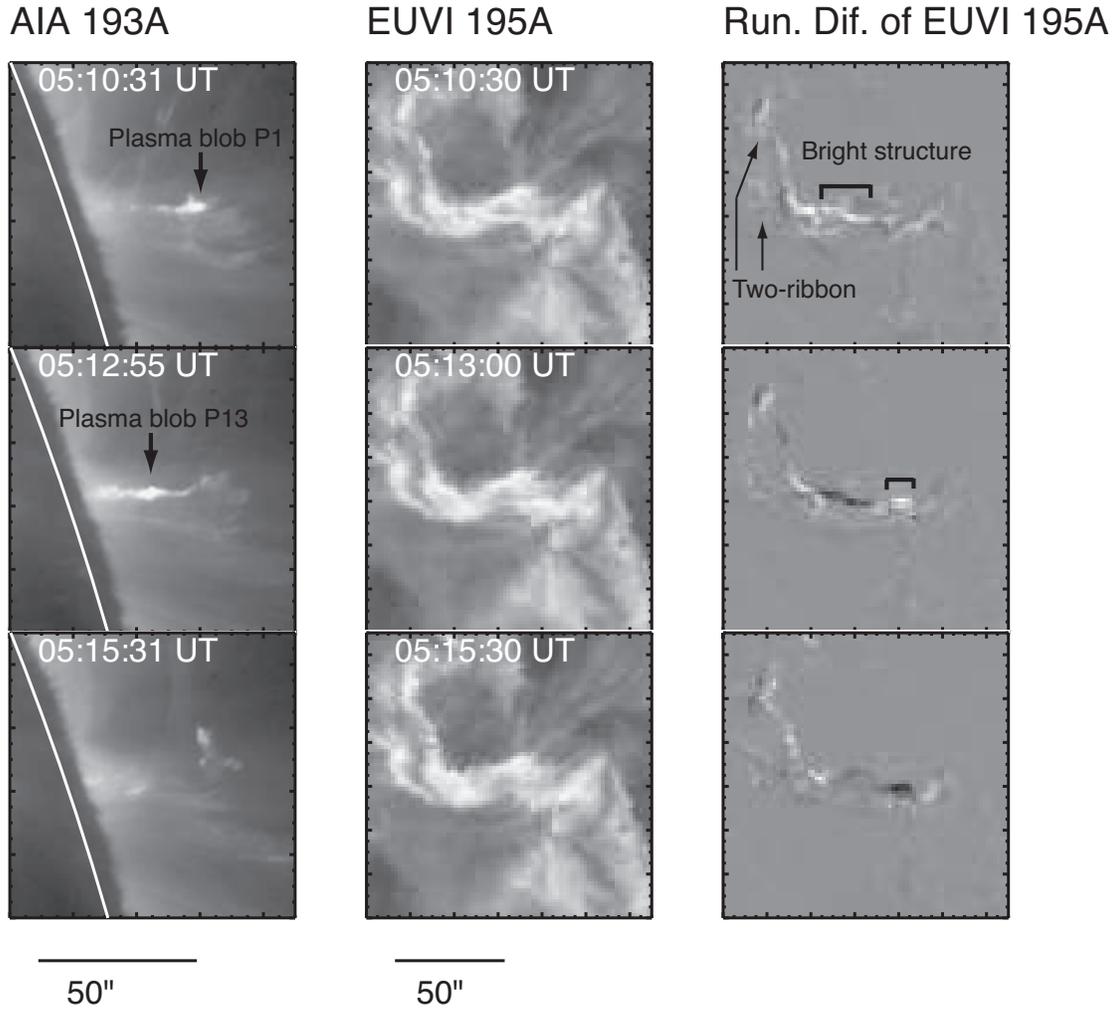}
\caption{Comparison between images of AIA 193~{\AA} and EUVI 195~{\AA} images. Running difference images of EUVI 195~{\AA} are also shown (the time difference between the successive images is 150~s). White solid lines in AIA images denote the solar limb. \label{fig:multi3d}}
\end{figure}

Figure~\ref{fig:multi3d} compares the temporal evolution of the flaring region seen by AIA and that by EUVI during the time when many plasmoids in the current sheet above the flare arcade were observed (05:10-05:14~UT). During this period, two EUVI 195~{\AA} images are available (05:10:30 and 05:13:00 UT) and can be compared with AIA images. When the upward ejection of the plasmoid P1 was observed by AIA, a horizontal straight structure possibly above the flare ribbons (05:10:30 UT) was detected in the EUVI 195~{\AA} image (see the running difference image of EUVI; the right column of Figure~\ref{fig:multi3d}). The direction of the elongation of the bright structure seen in EUVI 195~{\AA} images almost corresponds to the direction perpendicular to the plane of the AIA 193~{\AA} images. The length is approximately 20$^{\prime\prime}$. Note that the bright structure is not caused by CCD saturation. The location and timing of the brightening in the EUVI 195~{\AA} image coincide well with those of the bright plasmoids in the AIA images, which indicates that the plasmoids seen in the AIA images have structures extending in the direction perpendicular to the plane of the AIA images. When the plasmoid P13 was observed by AIA, another brightening with the size of $\sim10^{\prime\prime}$ was also detected by EUVI. When plasmoids disappeared in AIA images, we see no clear brightenings on the flare ribbons (see the third row of Figure~\ref{fig:multi3d}).

\section{SUMMARY AND DISCUSSION}
We have investigated the morphology and plasma dynamics of the magnetic reconnection region and the associated radio bursts in the limb flare on 2010 August 18. We examined in detail the relation between the nonthermal radio bursts and the plasma motions in the current sheet above the hot flare arcade. Many plasmoids appeared in the current sheet and collided with each other or were ejected from it. As reported by \citet{takasao2012}, the overall characteristics of this flare are consistent with the classical CSHKP model: prominence eruption, reconnection inflow and outflow, and hot flare arcade. For this reason, our findings could be helpful for understanding electron acceleration processes in many flares.

We found seven nonthermal radio bursts (B1, B2, ..., and B7) during the time when the current sheet is visible. We discovered a good match of the timing between the radio bursts and plasmoid motions. A match of the location is also found for the radio bursts except for the burst B1 and B6. The radio burst B1 followed the plasmoid ejection of P1. The burst B3 occurred at the timing of the coalescence of the plasmoids P2 and P3. The burst B4 was followed by the collision of P5 with the flare arcade, or by the coalescence of P6 and P7. The location and timing of the burst B5 coincide well with those of the collision of P9 with the flare arcade. The burst B6 occurred when three plasmoids P10, P11 and P12 collided or merged with each other. The last burst B7 was detected around the region where P13 collided with the flare arcade. The timing of the radio burst is also in good agreement with that of the collision. The radio sources of B1 and B6 were separated from the related plasmoids in spite of the match of the timing. We have no clear explanation for this separation, but the trapping of electrons accelerated around plasmoids in a different location may be one reason. The burst B2 was detected near the sunward edge of the current sheet. For this reason, B2 may be related to the sunward reconnection outflow. However, the current observations are insufficient to study the plasma motions related to this burst.

\begin{figure}
\epsscale{1.0}
\plotone{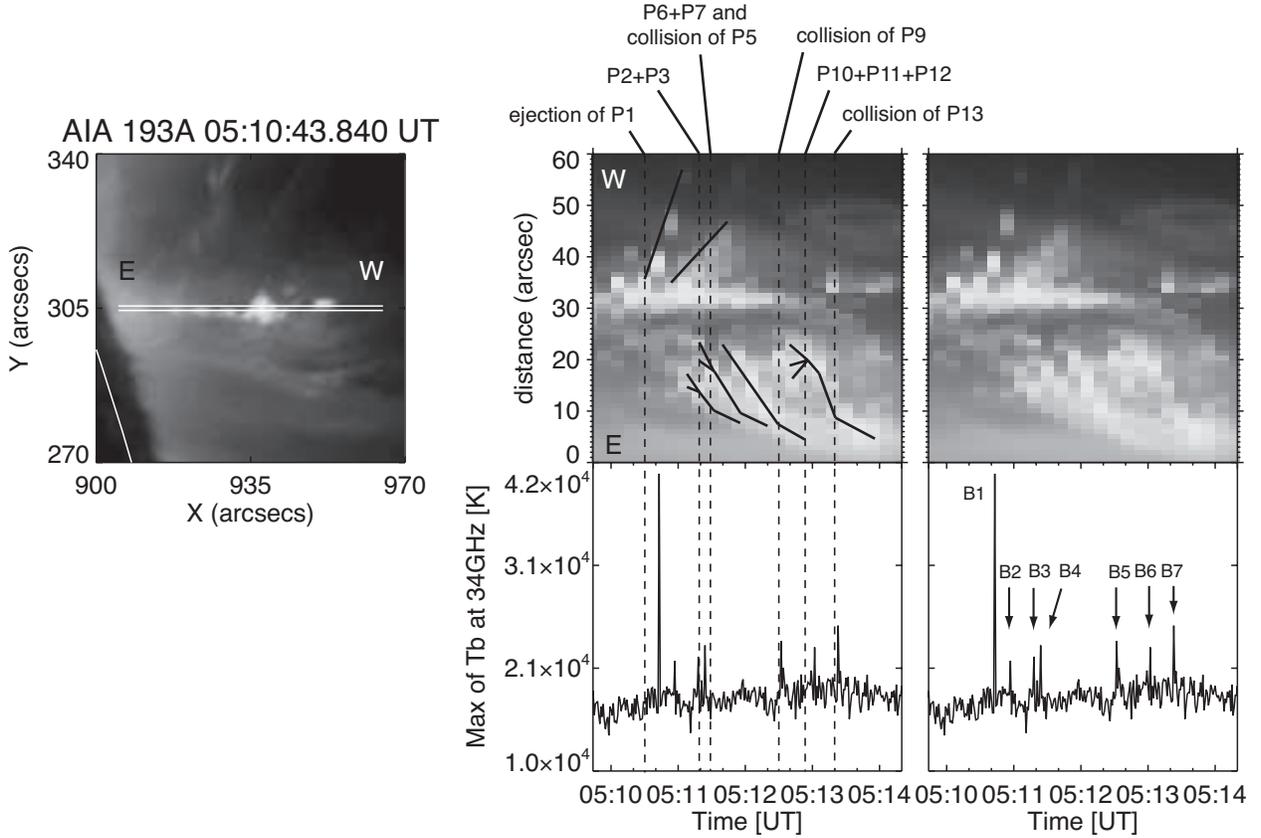}
\caption{Comparison between the plasma motions and radio bursts. Left: Snapshot of the flaring region taken by the AIA 193~{\AA} channel. Two slits EW (E: east, W: west) used to make the time-sequenced image in the right panel are also shown. Right: Time-sequenced images obtained along the slits (Top) and the temporal evolution of the maximum brightness temperature $T_b$ at 34~GHz (Bottom). The intensity shown is averaged over the two slits. The trajectories of plasmoids are indicated by solid lines. The radio bursts are labeled as B1, B2,... and B7 (B denotes ``burst").\label{fig:slit_burst}}
\end{figure}

We have no direct evidence for concluding whether plasmoids had indeed collided or remained separate entities from our observational data set. However, we can infer the plasmoid dynamics from the time-sequenced images (Figure~\ref{fig:slit_burst}). If plasmoids did not collide, they should pass by each other without significant deceleration, which will be recognized in the time-sequenced images. However, the figure shows that apparently colliding plasmoids were decelerated (this is evident for the collision between P10, 11, and 12). For this reason, we interpret that the plasmoids were indeed collided.

\begin{figure}
\epsscale{.80}
\plotone{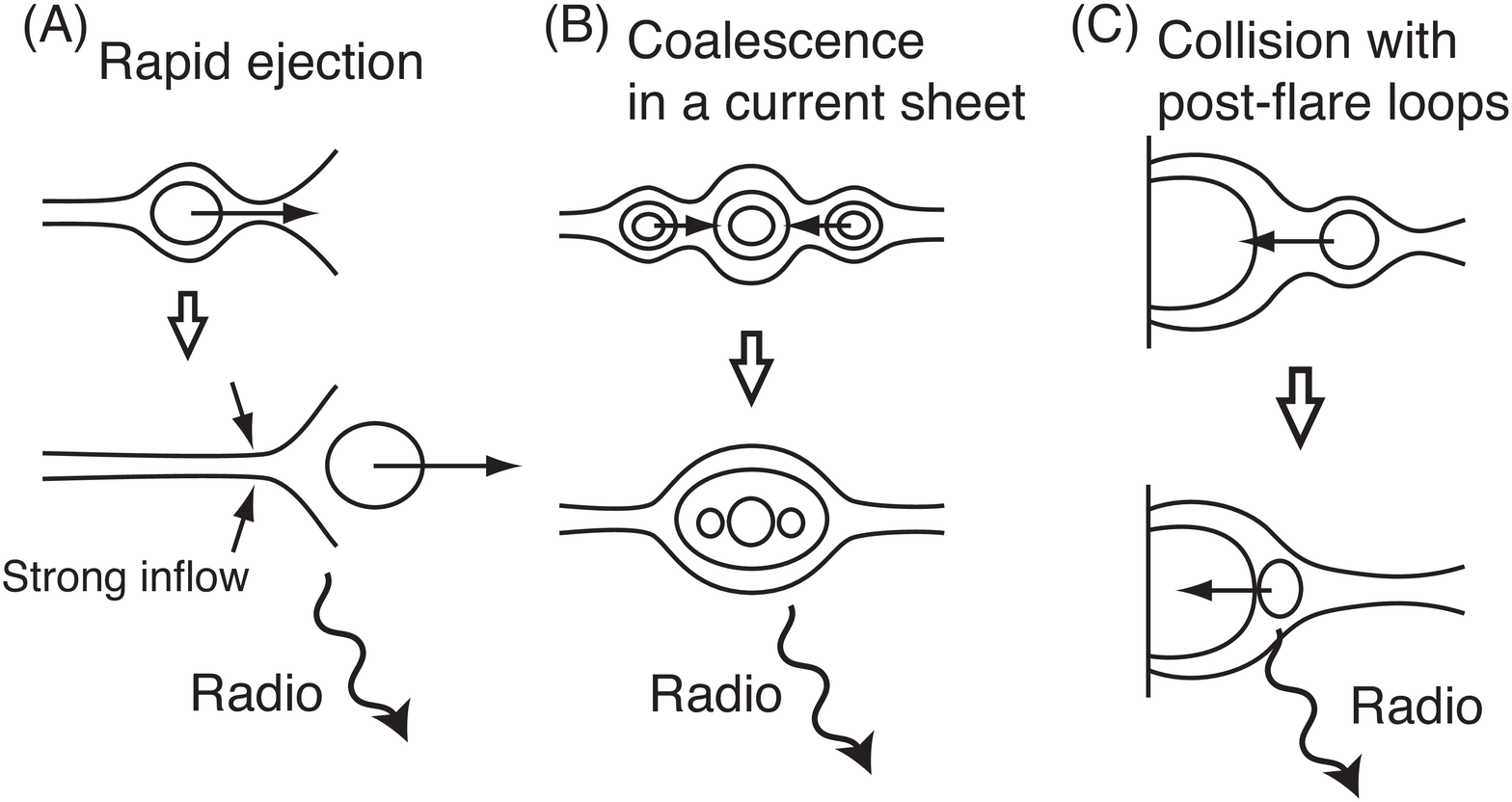}
\caption{List of possible plasmoid motions that lead to the observed radio burst events. We consider that the radio bursts are gyrosynchrotron radiation by nonthermal high energy electrons. \label{fig:list}}
\end{figure}

We summarize the list of possible plasmoid motions that can lead to the electron acceleration in Figure~\ref{fig:list}. The rapid ejection of a plasmoid induces a strong reconnection inflow, and impulsively raises the reconnection rate or reconnection electric field \citep[e.g.][]{nishida2009}. The radio burst B1 could be caused by the increase of the reconnection electric field. The observed collision or coalescence of plasmoids could be a result of the coalescence instability \citep{finn1977}. It has been argued that the coalescence leads to an efficient particle acceleration through a secondary reconnection at the merging points \citep{tajima1987,kliem1994,tanaka2010,oka2010} and through the contraction of merging plasmoids \citep{drake2006}. For this reason, we consider that the coalescence of plasmoids is a candidate for causing the radio bursts B3, B4, and B6. Since the flare arcade can be regarded as a large plasmoid, we expect that the same mechanisms can be applied to the situation in which plasmoids collided with the flare arcade. In addition, the Fermi-type acceleration in plasmoids interacting with termination shocks may be possible \citep{nishizuka2013}. We infer that this mechanism, as well as other acceleration mechanisms for the coalescence of plasmoids, are relevant to the observed radio bursts associated with the collision of plasmoids with the flare arcade (B4, B5, and B7).

The preheating of electrons before the acceleration is crucial for determining the number of the high energy electrons \citep[e.g.][]{tsuneta1998,nishizuka2013}. The current sheet containing plasmoids is recognized even in the AIA channels sensitive to hot plasma \citep[94 and 131~{\AA},][]{takasao2012}. Therefore, the plasmoids could contain hot plasma with the temperature of $10$~MK (equivalent to $\sim1$~keV) \citep[although caution is necessary because these channels are also sensitive to cooler plasma;][]{foster2011}. The current observations are insufficient to fully understand the heating process, but the coalescence of plasmoids we observed may be one possible heating process \citep[for numerical simulations of the coalescence, see e.g.][]{karlicky2011}. It is also possible that slow- and fast-mode shocks that have formed around plasmoids heat up the plasma \citep{tanuma2001,mei2012,shibayama2015,takeshige2015}. Our observations may provide an example in which plasmoid motions play important roles in heating the plasma and therefore accelerating electrons.

Stereoscopic observations enabled us to infer the 3D structure of two plasmoids P1 and P13. Considering the top view taken by EUVI 195~{\AA}, the bright plasmoids seen in the AIA images had a structure extending in the direction perpendicular to the plane of the AIA images. We consider that the plasmoids were flux ropes created in the current sheet. The plasmoid size seen in AIA images is approximately $\sim3^{\prime\prime}$, and the size seen in EUVI images is approximately 10--20$^{\prime\prime}$. From these results, the aspect ratio is estimated to be 0.1--0.3. The elongation of plasmoids may be due to the existence of a strong guide field \citep[e.g.][]{linton2003,wang2015}. This interpretation is supported by the fact that the plasmoids were observed only in the early phase of this flare during which the free magnetic energy is expected to be stored in the form of magnetic shear near and along the polarity inversion line.

This study strongly supports the idea of the electron acceleration associated with the plasmoid motions, but detailed acceleration processes remain puzzling. It is unclear if existing models can actually account for the observation because most of the models assume an idealized situation. The formation of plasmoids in the current sheet will introduce a large fluctuation to the reconnection regions and outflows \citep{tanuma2001}. This will affect the pictures assumed in some previous models like the \citet{nishizuka2013} model. A recent magnetohydrodynamics simulation by \citet{takasao2015} and \citet{takasao2016} revealed that the termination shock structure is more complex and dynamic than ever thought. Since the previous standard picture of a solar flare is being updated, it is important to reconsider previous models and develope more advanced models that consider the dynamic nature of magnetic reconnection.

\acknowledgements
We thank Dr. H. Nakajima for fruitful comments on the manuscript. S.T. acknowledges support by the Research Fellowship of the Japan Society for the Promotion of Science (JSPS).
This work was supported by JSPS KAKENHI grant numbers 25287039, 23340045, and 15K17772. We are grateful to SDO/AIA and STEREO/EUVI teams for providing the data used in this study. This work was performed using NoRH operated by the International Consortium for the Continued Operation of Nobeyama Radioheliograph (ICCON). ICCON consists of ISEE/Nagoya University, NAOC, KASI, NICT, and GSFC/NASA.

{\it Facilities:} \facility{SDO}, \facility{STEREO}, \facility{NoRH}

\end{document}